\documentstyle[12pt]{article}

\begin{document}

\newcommand{\ur}[1]{~(\ref{#1})~}
\newcommand{\urs}[2]{~(\ref{#1},~\ref{#2})}
\newcommand{\urss}[3]{~(ref{#1},~\ref{#2},~\ref{#3})}
\newcommand{\eq}[1]{eq.~(\ref{#1})}
\newcommand{\eqs}[2]{eqs.(\ref{#1},\ref{#2})}
\newcommand{\eqss}[3]{eqs.(\ref{#1},\ref{#2},ref{#3})}
\newcommand{\eqsss}[2]{eqs.(\ref{#1}--\ref{#2})}
\newcommand{\Eq}[1]{Eq.~(\ref{#1})}
\newcommand{\Eqs}[2]{Eqs.(\ref{#1},\ref{#2})}
\newcommand{\Eqsss}[2]{Eqs.(\ref{#1}--\ref{#2})}
\newcommand{\fig}[1]{Fig.~\ref{#1}}
\newcommand{\figs}[2]{Figs.\ref{#1},\ref{#2}}
\newcommand{\figss}[3]{Figs.\ref{#1},\ref{#2},\ref{#3}}
\newcommand{\beq}{\begin{equation}}
\newcommand{\eeq}{\end{equation}}
\newcommand{\e}{\varepsilon}
\newcommand{\ee}{\epsilon}
\newcommand{\la}[1]{\label{#1}}
\newcommand{\ui}{$U_{int}\,$}
\newcommand{\r}{{\bf r}}
\newcommand{\doublet}[3]{\: \left(\begin{array}{c} #1 \\#2
\end{array} \right)_{#3}}

\def\appendix{\par
 \setcounter{section}{0}
\def\thesection{Appendix}}
\vspace{2.5cm}
\begin{center} \Large
NON-PERTURBATIVE ISOTROPIC MULTI-PARTICLE PRODUCTION IN YANG--MILLS
THEORY \end{center}
\vspace{2cm}

\begin{center}
{\large Dmitri Diakonov}
\footnote{Alexander von Humboldt Forschungspreistr\"{a}ger}
{\large and Victor Petrov} \\
{\small\it St.Petersburg Nuclear Physics Institute, Gatchina, St.Petersburg
188350, Russia} \\ and\\ {\small\it Institut f\"ur Theor. Physik-II,
Ruhr-Universit\"at-Bochum, 44780, Bochum, B.R.D.}
\end{center}
\vspace{1cm}
\thispagestyle{empty}
\vspace{2cm}
\abstract{
We use singular Euclidean solutions to find multi-particle production
cross sections in field theories. We investigate a family of
time-dependent O(3) symmetrical solutions of the
Yang--Mills equations, which govern the isotropic high-energy gauge boson
production. At low energies our approach reproduces the instanton-induced
cross sections. For higher energies we get new results. In particular, we
show that the cross section for isotropic multiparticle production remains
exponentially small in the running gauge coupling constant. The result
applies both to the baryon number violation in the electro-weak theory and
to the QCD jet production.  We find that the isotropic multi-gluon
production cross section falls off approximately as a ninth power of
energy but possibly might be observable}

\newpage

\section{Introduction}
\setcounter{equation}{0}
\def\theequation{\arabic{section}.\arabic{equation}}

The last few years have witnessed an increasing interest in the
non-perturba\-tive multi-particle production induced by classical
solutions of the field equations. Typical examples are (i) baryon number
violation in the electro-weak theory (for reviews see
\cite{Mattis,Tinyakov,Ringwald},
(ii) multi-jet production in strong interactions \cite{Zakharov1,BB} and
(iii) multi-pion production in heavy ion collisions \cite{Anselm,BK,BD}.

While there seems to be a consensus in that classical field configurations
are relevant in $many\;\rightarrow\;many$ processes (see, e.g.
ref.\cite{KRT,Mue1}), it is not so clear in the case of the
$2\;\rightarrow\;many$ ones. The difficulty with the latter processes
is that an initial state with a few number of particles is a quantum rather
than a classical one -- even at asymptotically high energies. If one starts
from a classical field with large energy at $t=+\infty$ and
evolves it back in time according to the classical equations of motion
which preserve the field energy, one would end up with another field
configuration at $t=-\infty$ with the same energy. Meanwhile, at
$t=-\infty$ the whole energy has to belong not to the classical field
(which should be zero or, at best, contain only the wrong-frequency part
not corresponding to any real particles) but to the few-particle quantum
state.  Therefore there cannot exist any {\em continuous} classical field
configuration interpolating in time between $t=\pm\infty$.

A way to overcome this difficulty has been indicated by Khlebnikov
\cite{Khlebnikov} who, following a remarkable work of Iordansky and
Pitaevsky \cite{IP}, has suggested to study {\em singular} Euclidean
solutions of field equations. Singular fields do not conserve
energy across singularities, therefore they are adequate to describe
transitions between a quantum state with low or zero field energy and a
(semi)classical one with high energy. Singular trajectories in imaginary
(Euclidean) time have been introduced 60 years ago by Landau to calculate
matrix elements between low- and high-energy states in quantum mechanics
\cite{LL}. Even if a final high-energy state can be described
semiclassicaly, the initial low-energy state needs not; therefore the Landau
theory gives an example how, nevertheless, the matrix elements can be
treated semiclassically. It should be mentioned that an application of the
Landau theory to multi-particle production has been suggested by
Voloshin \cite{Vol}, however only constant fields and hence
relatively low energies have been considered in that work.

The energy of the initial state, i.e. the collision energy,
is introduced as follows: Let the multiparticle production be
initiated by, say, an annihilation of two high-energy particles with
energies $E_{1,2}$ in the c.m. frame. To get the physical on-shell
amplitude one has to calculate the two-point Green function in the
singular background field, take its Fourier transform and then apply
the leg amputation procedure of Lehmann--Symanczik--Zimmermann.
The high-energy asymptotics of the Green function is apparently
given by the Fourier transform of the singularity point \cite{IP}, which is
$exp\left(-i(E_1+E_2)t_{sing}\right)$ in Minkowski space, or, if one
passes to the Euclidean space with the usual substitution $-it\rightarrow
t$, one gets $exp(Et_{sing})$ where $E=E_1+E_2$ is the total c.m. energy
of the process and $t_{sing}$ is the time position of the field singularity
in Euclidean space. We will see below that integrating over the positions
of the singularity will result in the necessary conservation law:
$E=E_{field}$ where $E_{field}$ is the energy of the produced multi-particle
state.

Our aim is to calculate semi-classically the total cross section induced
by two high-energy particles with a total energy $E$, i.e. the imaginary
part of a forward scattering amplitude which, in its turn, can be
expressed through the 4-point Green function (we use $\phi({\bf r},t)$ to
describe a generic field):

\[
\sigma(E^2=(p_1+p_2)^2) = ({\rm flux\;factor})\;
\lim_{p_{1,2}^2\rightarrow m^2} {\mbox Im}\int d^4x_{1-4}
e^{-ip_1\cdot (x_1-x_3)-ip_2\cdot (x_2-x_4)}
\]
\beq
\left\langle(\partial_1^2+m^2)\phi(x_1)(\partial_2^2+m^2)\phi(x_2)
(\partial_3^2+m^2)\phi(x_3)(\partial_4^2+m^2)\phi(x_4)\right\rangle.
\la{sigma}\eeq

The purpose of the paper is to demonstrate how to calculate the
multi-particle production cross sections given generically by \eq{sigma},
using the technique of singular Euclidean solutions.

\section{Singular Trajectories in Quantum Mechanics}
\setcounter{equation}{0}
\def\theequation{\arabic{section}.\arabic{equation}}

Before proceeding to the Yang--Mills case we describe a quantum mechanical
analogue of our approach. We consider a particle with coordinate
$\phi(t)$ in a double-well potential, whose (Euclidean) action is given by

\beq
S=\frac{1}{g^2}\int dt \left[\frac{1}{2}\dot{\phi}^2
+\frac{1}{8}(\phi^2-1)^2\right],
\la{dwa}\eeq
where $g^2\ll 1$ is a dimensionless coupling constant.

\subsection{Quantum-Mechanical "Cross Section"}

In the quantum-mechanical case "mass shell" means fixed energy for
each "particle", equal to 1 in our notations. Therefore, to get an analogue
of high-energy cross section one has to consider an "off-mass-shell"
amplitude. Since we are only interested in the cross section computed to an
exponential accuracy, it does not matter what specific operator $O(\phi)$
do we use to go off-shell.  We choose $O(\phi)$ to be a low-power
polynomial of $\phi$.  The analogue of \eq{sigma} in quantum mechanics
would be

\beq
\sigma(E)={\mbox Im} \int dt_{1,2} e^{-iE(t_1-t_2)}
\left\langle O(\phi(t_1)) O(\phi(t_2))\right\rangle.
\la{sigmaqm}\eeq

One can decompose the Green function $\langle ...\rangle$ as a sum over
intermediate states with wave functions $\Psi_n(\phi)$:

\[
\langle O(\phi(t_1)) O(\phi(t_2))\rangle =
\sum_n \left|\int d\phi \Psi_0(\phi) O(\phi) \Psi_n^\star(\phi)\right|^2
\]
\beq
\cdot\left[\theta(t_1-t_2)e^{-iE_n(t_1-t_2)}
+\theta(t_2-t_1)e^{-iE_n(t_2-t_1)}\right].
\la{intermed}\eeq
Integrating over $t_{1,2}$ in \eq{sigmaqm} and taking the imaginary part we
get

\beq
\sigma(E)=\pi\delta(E_n-E)V
\left|\int d\phi \Psi_0(\phi) O(\phi) \Psi_E^\star(\phi)\right|^2
\la{mael}\eeq
where $V$ is the time volume of the process (actually to be cancelled
by the omitted "flux factor" in \eq{sigmaqm}). We see thus that the
quantum-mechanical "cross section" comes to a square of a matrix element
between the ground state $\Psi_0$ and a highly excited one $\Psi_E$.

According to the Landau theory \cite{LL} this matrix element can be
calculated to the exponential accuracy as an exponent of the difference of
two shortened actions, one with zero energy and the other with the given
energy $E$:

\beq
\sigma(E) = \exp 2\left[- \int d\phi \sqrt{2U(\phi)}+
\int d\phi \sqrt{2U(\phi)-2E}\right]
\la{Landau}\eeq
where $U(\phi)$ is the potential energy of the field.

\Eq{Landau} is directly applicable to the simple case of a one-minimum
potential; in this case one has to integrate in the first term from
$\phi_{min}$ corresponding to the minimum of the potential, to
$\phi=\infty$, and in the second term from the energy-dependent turning
point $\phi_1$ where the integrand nullifies, to $\phi=\infty$. For the
square of the amplitude one has to double the result in the exponent.

We arrive thus to singular Euclidean trajectories:
one starts at (Euclidean) time $t=-\infty$ from the minimum of the
potential, goes according to the equations of motion with zero energy to
infinity (singularity) at some $t=-T/2$, then, starting from the same
singularity, goes with fixed energy $E$ to the corresponding turning point
$\phi_1$ at $t=0$. At the turning point $\dot{\phi}=0$, and one can
enter the Minkowski space to calculate the real-time evolution of the
field.  However, the Minkowski action is a pure phase, therefore to find
the cross section one needs not know this part of the trajectory. Instead,
one has to square the amplitude, i.e. to repeat the singular Euclidean
trajectory in the opposite direction: starting from the above turning point
$\phi_1$ at $t=0$, going with energy $E$ to a singularity at $t=+T/2$, and
ultimately returning to the minimum of the potential at $t\rightarrow +
\infty$. The different branches of the trajectory are shown schematically
in Fig.1a.

The Yang--Mills theory is however more like the double-well quantum
mechanics, therefore we have to generalize \eq{Landau} to the case of
two-minimum potential as in \eq{dwa}, allowing for the instanton
transitions between the minima. To that end we have first of all to specify
the initial and final states $\Psi_{0,E}$ entering \eq{mael}. In a
double-well potential all levels are known to be split into symmetric ($s$)
and antisymmetric ($a$) states. For levels deep inside the wells the
corresponding wave functions can be written as superpositions of states
localized in the left and in the right wells:

\[
\Psi_{s,a}(\phi)= \frac{\Psi(\phi+1)\pm\Psi(\phi-1)}{\surd 2},
\]
\beq
\Psi(\phi\pm 1)= \frac{\Psi_s(\phi)\pm\Psi_a(\phi)}{\surd 2}.
\la{superpos}\eeq

To mimic in quantum mechanics the non-perturbative Chern--Simons changing
transitions of the Yang--Mills theory we choose the initial state with
low energy ($\Psi_0$ in the notations of \eq{mael}) to be localized near
the left minimum and having approximately zero energy,

\beq
\Psi_0(\phi)=\frac{\Psi_s^{\approx 0}(\phi)+\Psi_a^{\approx 0}(\phi)}
{\surd 2}.
\la{left}\eeq
For the high-energy state $\Psi_E$ we take a superposition which is
predominantly localized in the right well and has energy approximately
equal to $E$:

\beq
\Psi_E(\phi)=\frac{\Psi_s^{\approx E}(\phi)-\Psi_a^{\approx E}(\phi)}
{\surd 2}.
\la{right}\eeq

Of course at energies $E$ higher than the top of the potential barrier
one cannot say anymore that the state \ur{right} is localized near the
right minimum, but it is probably the best one can do to define
continuously in energy the transitions between the left and right wells.
In the Yang--Mills theory a much more clear trigger of the
transition with the change of the Chern--Simons number is given by the
accompanying change of fermion chirality or baryon/lepton number
violation. However, for energies
less than the barrier top the state $\Psi_E$ \ur{right} is indeed
localized near the right minimum. For energies higher than the barrier
matrix elements for the transition to any state are anyhow exponentially
small, so the concrete definition of the final state does not make a great
difference.

With this definition of the initial and final states, \eq{Landau} has to be
slightly modified. For energies higher than the potential barrier the
modification is cosmetic: the singular trajectory described above has to
start in the specific minimum $\phi=-1$; it ends up in the same minimum --
see Fig.1a. For energies lower than the barrier the Minkowskian part of the
trajectory starting from the turning point $\phi_1$ (see Fig.1b) hits
another turning point $\phi_3$ where it again enters the forbidden zone
and hence develops in imaginary (Euclidean) time. This branch (V) is a
bounce resembling the kink plus anti-kink: at $t=0$ (chosen for symmetry
reasons) the trajectory reaches the turning point $\phi_4$ belonging rather
to the right well. From this point one can proceed in Minkowski time
observing the final-state field in the right well. However, the Minkowski
action, being a pure phase, does not contribute to the cross section;
in order to obtain it one has to repeat all the trajectory in the opposite
order and to count the actions (with appropriate signs) along the branches
I-V -- see Fig.1b. We get for the actions along the different branches:

\[
S^I=S^{IV}=\frac{1}{2g^2}\int_{-\infty}^{-1}d\phi (\phi^2-1),
\]
\[
S^{II}=S^{III}= \frac{1}{2g^2}\int_{-\infty}^{\phi_1=-\sqrt{1+\surd
\e}}d\phi \sqrt{(\phi^2-1)^2-\e},
\]
\beq
S^V= \frac{1}{g^2}\int_{\phi_3=-\sqrt{1-\surd \e}}^{\phi_4=\sqrt{1-\surd \e}}
d\phi \sqrt{(\phi^2-1)^2-\e},
\la{actions}\eeq
where we have introduced a dimensionless energy $\e=8g^2E$.
The final formula for the cross section is

\beq
\sigma(E)=\exp\left(-S^I+S^{II}-S^V+S^{III}-S^{IV}\right).
\la{5terms}\eeq
At energies higher than the barrier ($\e>1$) the branch V is absent, and
there is no contribution of $S_V$ to $\sigma(E)$ so that we return to
\eq{Landau}.

Let us comment on a few features of \eqs{actions}{5terms}. The
actions $S^I$ to $S^{IV}$ each diverge at $\phi=-\infty$ however their sum
is finite.  To see that explicitly we can rewrite their sum as

\[
S^{I-IV}\equiv S^I +S^{IV}-S^{II}-S^{III}
\]
\beq
= \frac{1}{g^2}\left\{(\phi_1+1)-\frac{\phi_1^3+1}{3}+
\int_{-\infty}^{\phi_1} d\phi\left[(\phi^2-1)-\sqrt{(\phi^2-1)^2-\e}
\right]\right\}\;>\,0,
\la{SI-IV}\eeq
which apparently is convergent. At $\e\rightarrow 0$ these branches cancel
altogether, and one is left with the piece $S^V$ which in this limit is
an action on the infinitely separated instanton (kink) and anti-instanton
(anti-kink), equal to $4/3g^2$. It gives the familiar Gamov suppression
factor for tunneling at zero energy. We thus rewrite \eq{5terms} as

\beq
\sigma(E)=\exp\left[-S^{I-IV}(E)-S^V(E)\right].
\la{2terms}\eeq

As the energy rises the behaviour of the cross section is determined by the
interplay of two opposite trends. The tunnelling probability,
represented by $\exp(-S^V)$, increases with energy whereas the overlap of
the initial low-energy state with the high-energy one, represented by
$\exp(-S^{I-IV})$, decreases. This physical picture has been suggested
several years ago in ref.\cite{BFDKS}; now we are in a position to make it
fully quantitative. At energies higher than the barrier there is no
compensation for that trend from the side of $S^V$, so the cross section
will definitely decrease, and we thus expect a maximum of the cross section
at energies of the order of the top of the barrier, $\e=1$.

\subsection{Calculation of the Cross Section}

The time dependence of the trajectories is given by

\beq
\frac{t-t_0}{2}=\int\frac{d\phi}{\sqrt{(\phi^2-1)^2-\e}};
\la{timedep}\eeq

For $\e=0$ (branches I and IV) we get trajectories going from $-1$ to
$-\infty$ at $t=\pm T/2$:

\beq
\phi^I(t)= {\rm cth}\frac{1}{2}(t+\frac{T}{2}),\;\;\;\;
\phi^{IV}(t)=-{\rm cth}\frac{1}{2}(t-\frac{T}{2})\;=\;\phi^I(-t).
\la{timedepI}\eeq
At $\e\neq 0$ (branches II, III and V) the trajectories are given by
elliptic functions of the first and second kind.

Actually in order to calculate the cross section \ur{5terms} as a function
of energy it is not necessary to know the trajectories explicitly: the
shortenned actions contain less information.  We first calculate the
derivatives of shortenned actions in respect to $\e$.  At $\e<1$ we have
({\bf K}(k) is the complete elliptic integral of the first kind \cite{GR}):

\[
g^2\frac{d\ln\sigma}{d\e}=-g^2\frac{dS^{I-IV}}{d\e}-g^2\frac{dS^V}{d\e}
\]
\[
=-\frac{1}{2}\frac{{\bf K}(k)}{\sqrt{1+\surd\e}}
+\frac{{\bf K}(k)}{\sqrt{1+\surd\e}}\;\;\;
\left(k=\sqrt{\frac{1-\surd\e}{1+\surd\e}}\right)
\]
\beq
=\frac{1}{8}\ln\frac{64}{\e}+\e\frac{3\ln\frac{64}{\e}-10}{128}
+O(\e^2).
\la{derivl}\eeq

At $\e>1$ we get:

\[
g^2\frac{d\ln\sigma}{d\e}=-g^2\frac{dS^{I-IV}}{d\e}=
-\frac{1}{2}\frac{{\bf K}(l)}{\sqrt{2\surd\e}}\;\;
\left(l=\sqrt{\frac{\surd\e-1}{2\surd{\e}}}\right)
\]
\beq
=-\frac{1}{8\e^{1/4}}B\left(\frac{1}{4},\frac{1}{2}\right)+O(\e^{-3/4}),
\;\;\;B\left(\frac{1}{4},\frac{1}{2}\right)=
\frac{\Gamma(\frac{1}{4})\Gamma(\frac{1}{2})}{\Gamma(\frac{3}{4})}.
\la{derivg}\eeq

Integrating these equations we get the cross section. At small energies
we obtain a rising cross section,

\beq
g^2\ln\sigma(\e)=-\frac{4}{3}+\frac{\e}{8}\left(\ln\frac{64}{\e}+1\right)
+\frac{\e^2}{256}\left(3\ln\frac{64}{e}-\frac{17}{2}\right)+O(\e^3),
\la{csl}\eeq
while at large energies we get a decreasing cross section:

\beq
g^2\ln\sigma(\e)=-\frac{\e^{3/4}}{6}B\left(\frac{1}{4},\frac{1}{2}\right)
+O(\e^{1/4}).
\la{csg}\eeq

The maximum of the cross section is achieved at $\e=1$ corresponding to the
top of the barrier, and the $\log$ of the cross section there is exactly
twice less than at $\e\rightarrow 0$ where the suppression is solely due to
the double instanton action:

\beq
g^2\ln\sigma_{max}=-\frac{2}{3}, \;\;\;\e=1.
\la{csmax}\eeq

\subsection{"Square Root" Suppression}

The "square root" suppression of the cross sections at the maximum as
compared to that at zero energies have been
advocated by Zakharov \cite{Zakharov2} and Maggiore and
Shifman \cite{MS} from unitarity considerations (see also ref.\cite{V}). It
is remarkable that we get it here in a different approach.  Probably it
means that the correct formulae respect unitarity.  Moreover, we can show
that this square root suppression is of a more general nature. Indeed, let
us compare $dS/d\e$ along the branches II,III and V for $\e<1$. According
to a general theorem of the classical mechanics (see Appendix A) these
derivatives are related to the periods of motion along these branches:

\[
8g^2\frac{dS^{I-IV}}{d\e}=
4\int_{-\infty}^{\phi_1}\frac{d\phi}{\sqrt{U(\phi)-\e}}=T_2,\;\;\;\;
-8g^2\frac{dS^{V}}{d\e}=
4\int_{\phi_3}^{\phi_4}\frac{d\phi}{\sqrt{U(\phi)-\e}}= T_1,
\]
\beq
U(\phi)=(\phi^2-1)^2,\;\;\;T=T_1+T_2.
\la{periods}
\eeq
Here $T_1$,$T_2$ are periods of motion along the branches $V$
and $II+III$ respectively. We are going to prove that $T_1 =2T_2$ for a
wide class of potentials $U(\phi)$.

To that end let us consider the difference of integrals along two
closed contours $\Gamma_1$ and $\Gamma_2$ in the complex $\phi$ plane (see
Fig.2):

\beq
2\int_{\Gamma_1}\frac{d\phi}{\sqrt{U(\phi)-\e}}-
2\int_{\Gamma_2}\frac{d\phi}{\sqrt{U(\phi)-\e}}=0
\la{cint}\eeq

This difference is zero as far as the potential energy $U(\phi)$ has no
singularities in the whole complex $\phi$ plane, so that the integrand
has only cuts at the turning points $\phi_{1-4}$. According to \eq{periods}
the integral along the contour $\Gamma_1$ is equal to $T_1$ while that
along $\Gamma_2$ is equal to twice $T_2$, as there are two equal
contributions from two cuts to that integral. Therefore we
indeed get the relation

\beq
T_1=2T_2; \;\;\;\;\;-\frac{dS^{V}}{d\e}=2\frac{dS^{I-IV}}{d\e}.
\la{relper}\eeq

Integrating this equation in $\e$ we arrive to the following
important relation first noticed in ref. \cite{Kisel}:

\beq
S^{V}(\e)=S^{V}(0)-2S^{I-IV}(\e).
\la{V-I}\eeq

\Eqs{relper}{V-I} are of a very general nature and are valid for any
double-well potentials without singularities in the complex plane. (For the
concrete potential considered \eq{relper} follows
directly from \eq{derivl}). Moreover, we think that \eqs{relper}{V-I} are
valid also in field theories where tunneling may occur, Y--M and Y--M--H
theories being an example. The analogue of \eq{cint} in field theory would
be the following contour integral in the complex plane of an arbitrary
parameter $s$ parametrising the solutions of eqs. of motion with a given
field energy, $\phi_E(x, s)$:

\beq
\int_\Gamma ds \frac{\sqrt{\int
d^3x\left(d\phi_E(x,s)/ds\right)^2}} {\sqrt{\int
d^3x\left\{\frac{1}{2}\left(\nabla_i\phi_E\right)^2+U[\phi] \right\}-E}}.
\la{FT}\eeq

One can choose the time $t$ as the parameter $s$, but that is not necessary
since \ur{FT} is re-parametrization--invariant. One can try to use
this freedom to introduce a parametrization $\phi_E(x,s)$ such as to ensure
that both the numerator and the denominator have no singularities in the
complex $s$ plane, except the cuts due to the turning points of
the denominator. (In the one-degree-of-freedom case such parameter $s$
is the "field" $\phi$ itself). If that goal is achieved, one would prove
the relations \urs{relper}{V-I} in field theory. We have not proven them
in a general case, however we confirm these relations for the
Y--M theory by a direct calculation in the next section -- at least for
small field energies.

At $\e=1$ one reaches the top of the barrier where $S_V$ vanishes.
According to \eq{V-I} it means that

\beq
S^{I-IV}(1)=\frac{S^{V}(0)}{2}=S^{inst}=\frac{1}{g^2}\frac{2}{3},
\la{maxcs}\eeq
which reproduces \eq{csmax} and gives the maximum of the cross section
corresponding to the square root suppresion as compared to that at zero
energy.

Combining \eqs{V-I}{maxcs} we can write:

\beq
\ln\sigma(\e)=\left\{\begin{array}{c} -2S^{inst}+S^{I-IV}(\e),
\;\;\;\e<1,\\ -S^{inst},\;\;\;\e=1,\\ -S^{I-IV}(\e),\;\;\;\e>1,
\end{array}\right.,\;\;\;S^{I-IV}(\e)>0.
\la{allen}\eeq

This equation shows a jump of the derivative at $\e=1$. It should be
mentioned though that in a parametrically narrow strip near $\e=1$
the semi-classical formulae are not valid anymore, and one would expect
a smooth match between the two regimes of \eq{allen}.

\subsection{Instanton Interactions}

Let us return to our concrete quantum mechanical example.
To make contact with the previous work on the instanton-induced
cross sections at low energies we recall that a conventional way to present
the results is via the instanton--anti-instanton interaction potential,
\ui. The cross section can be written as a Fourier transform of the
time separation of the instanton and the anti-instanton
\cite{Zakharov3,KR,WS,DPol}:

\beq
\sigma(E)=\int dT \exp\left[ET-U_{int}(T)\right].
\la{uint}\eeq
We define here \ui so that at large
separations $T$ it includes also twice the free instanton action.
The integral over $T$ is performed by saddle point method -- that is how
one gets the energy dependence of the cross section. Comparing \eq{uint}
with $\sigma(E)$ found above (\eq{csl}) we obtain the instanton interaction
potential at large separations:

\beq
U_{int}(T)=\frac{1}{g^2}\left[\frac{4}{3}-8e^{-T}-
48Te^{-2T}+136e^{-2T} + O(e^{-3T})\right].
\la{uintqm}\eeq

The first term here is twice the instanton action, the second one
reproduces correctly the well-known kink--anti-kink leading-order
attraction (see, e.g., ref.\cite{ZJ}). The third term is new: we can
compare it only with the valley approach of Balitsky and Yung
\cite{BY,KR,Ver}. For the sum ansatz of a kink and anti-kink these authors
get

\beq
U_{int}^{valley}(T)=\frac{1}{g^2}\left[\frac{4}{3}-8e^{-T}+
24Te^{-2T}-40e^{-2T} + O(e^{-3T})\right],
\la{uintval}\eeq
which coincides with our result only in the leading order. To our
mind, it means only that the sum of a kink and anti-kink is not
too useful beyond the leading order. In an indirect way, however, our
result for the next-to-leading kink--anti-kink interaction seems to be in
accordance with another calculation, this time for the
instanton--anti-instanton interaction in the Y--M theory. Indeed, as
emphasized in refs.\cite{BY,KR,Ver}, an instanton--anti-instanton
confuguration can be obtained from a kink--anti-kink one through a
chain of conformal and gauge transformations. The separation between
the Y--M instantons $R$ and their sizes $\rho_{1,2}$ are related to
the separation of a kink and anti-kink $T$ as follows,

\beq
e^{T/2}=\frac{R^2+\rho_1^2+\rho_2^2+\sqrt{(R^2+\rho_1^2+\rho_2^2)^2-
4\rho_1^2\rho_2^2}}{2\rho_1\rho_2},
\la{BYKRVer}\eeq
while the coupling $g^2$ should be replaced by the gauge coupling $\alpha$
according to the rule

\beq
\frac{1}{g^2}\frac{4}{3}\;\rightarrow\;\frac{4\pi}{\alpha}.
\la{couplings}\eeq

Therefore, the kink--anti-kink interaction \ur{uintqm} can be transformed
into the instanton--anti-instanton interaction. We get from
\eqs{uintqm}{BYKRVer}:

\[
U_{int}^{Y-M}(R)=\frac{4\pi}{\alpha}\left[1-6\frac{\rho_1^2\rho_2^2}{R^4}
+12\frac{\rho_1^2\rho_2^2(\rho_1^2+\rho_2^2)}{R^6}\right.
\]
\beq
-\left.72\frac{\rho_1^4\rho_2^4}{R^8}\log\frac{R^2}{\rho^2}+
O\left(\frac{\rho^8}{R^8}\right)\right].
\la{uintym}\eeq

These terms are {\em exactly} those which are independently known today for
the Y--M instanton interactions. It should be stressed that the last term
has been computed from unitarity in a laborous two-loop calculation in the
Y--M theory \cite{DPol} \footnote {Recently the result has been confirmed
in a non-less laborous calculation in ref.\cite{BS}) for the case of the
maximum-attraction orientation, though there is still a difference between
refs. \cite{BS} and \cite{DPol} for general orientations.}. It is remarkable
that we reproduce the result of these hard calculations in such a simple
way.

\section{Singular Yang--Mills Fields}
\setcounter{equation}{0}
\def\theequation{\arabic{section}.\arabic{equation}}

We are now turning to the Y--M theory and are going to describe the
analogues of the branches I-V (see Fig.1), assuming that the cross section
of the $2\rightarrow many$ processes are to the exponential accuracy
given by \eq{5terms}. A derivation of the generalization of the Landau
formula to field theory using the first order formalism, can be found in
the paper of Iordansky and Pitaevsky \cite{IP}. We shall reproduce
the well-known results for the instanton-induced cross sections at
relatively low energies and get new results for high energies.
In the case of QCD we shall get the cross section of isotropical
multi-gluon production as function of energy: it decreases as approximately
the ninth power of energy.

\subsection{Exact Singular Solutions with Zero Energy}

In principle one could consider various types of field singularities but in
this paper we restrict ourselves to the case which at low energies
reproduces the instanton-induced processes. These processes are known to
posses the O(3) symmetry, meaning that the multiparticle production is
spherically-symmetri\-cal. One might well doubt whether it is natural for
high-energy collisions \footnote {We a grateful to C.Wetterich for a
discussion of this point.} and in a sense we prove that it is not, but at
the moment we would like to follow the instanton-induced processes up to
asymptotically high energies. For that reason we choose the singularity
to be O(3) symmetrical, meaning that at given times $\mp T/2$ the $A_\mu$
field has a power-like singularity in $r$ where $r$ is the distance from
the origin.

For branches I and IV corresponding to $E=0$ exact solutions of the
needed type have been already constructed by Khlebnikov \cite{Khlebnikov}:
They are the usual BPST instanton (branch I) and antiinstanton (branch IV)
in the singular Lorentz gauge with the scale parameter $\rho^2$ changed to
$-\rho^2$:

\[
A_\mu^{(I)a}(\r,t)=\frac{2\bar{\eta}_{\mu\nu}^a
\doublet{\r}{t+\frac{T}{2}-\rho}{\nu}\rho^2}
{\left[\rho^2-r^2-\left(t+\frac{T}{2}-\rho\right)^2\right]
\left[r^2+\left(t+\frac{T}{2}-\rho\right)^2\right]},
\]
\beq
t<-\frac{T}{2},\;\;\;\;{\rm branch\;I},
\la{I}\eeq
\[
A_\mu^{(IV)a}(\r,t)=\frac{2\eta_{\mu\nu}^a
\doublet{\r}{t-\frac{T}{2}+\rho}{\nu}\rho^2}
{\left[\rho^2-r^2-\left(t-\frac{T}{2}+\rho\right)^2\right]
\left[r^2+\left(t-\frac{T}{2}+\rho\right)^2\right]},
\]
\beq
t>\frac{T}{2},\;\;\;\;{\rm branch\;IV},
\la{IV}\eeq
where $\eta,\bar{\eta}$ are 't Hooft symbols \cite{tH}.

\Eq{I} describes the evolution of the field with zero energy starting from
zero field at $t=-\infty$ and getting to a singularity at $t=-T/2$. It is
the analogue of \eq{timedepI} in the quantum mechanical example. \Eq{IV}
describes the conjugate process: it goes from a singularity at $t=T/2$ to
zero field at $t=+\infty$. In case of the EW theory with Higgses one can
neglect the influence of the Higgses as far as the characteristic scale of
the fields is $\rho \ll m_W^{-1}$ where $m_W$ is the W boson mass. This is
the case in two limits we are now interested in: at $E\ll m_W/\alpha$ where
we shall reproduce the well-known result, and at $E\gg m_W/\alpha$, see
below.  In the intermediate energy region one would have to solve the
coupled Y--M--H system, otherwise \eqs{I}{IV} are exact.  At $t=\mp T/2$
these solutions become singular at the origin, $r=0$:

\beq
A_\mu^{(I)a}\left(\r,-\frac{T}{2}\right)=
\frac{2\bar{\eta}_{\mu\nu}^a\doublet{\r}{-\rho}{\nu}}{-r^2(r^2+\rho^2)},
\la{singI}\eeq
\beq
A_\mu^{(V)a}\left(\r,\frac{T}{2}\right)=
\frac{2\eta_{\mu\nu}^a\doublet{\r}{\rho}{\nu}}
{-r^2(r^2+\rho^2)}.
\la{singIV}\eeq

We notice that \ur{singI} and \ur{singIV} coincide for spatial components
$A_i$ but differ in sign for the time component $A_4$. This is a gauge
artifact, however. Indeed, one can perform a time-dependent gauge
transformation

\beq
A_\mu \rightarrow U^\dagger(\r,t) A_\mu U(\r,t)
+iU^\dagger\partial_\mu U
\la{GT}\eeq
eliminating the $A_4$ components in both branches, I and IV. (In fact
$A_4=0$ is the adequate gauge to use: all results are physically more
transparent in it; we started with a Lorentz gauge to make formulae more
compact). It is easy to check that making a gauge transformation \ur{GT}
we do not destroy the coincidence of the spatial components $A_i$ at
the singularity points. We cite the neccessary formulae in Appendix
B. The time-dependent gauge transformations \ur{GT} are however defined up
to time-independent transformations. In order not to introduce new
quantities, we use this gauge freedom to keep the $A_i$ fields at
$t=\pm T/2$ exactly in the form given by \eqs{singI}{singIV}:

\beq
A_i^{(I)\,a}\left(\r,-\frac{T}{2}\right)=
A_i^{(IV)\,a}\left(\r,\frac{T}{2}\right)=
-\frac{2\rho^2}{r^2(r^2+\rho^2)}
\left(\ee_{aij}r_j+\delta_{ai}\rho\right),\;\;\;A_4^a=0.
\la{boundary}\eeq

In general, in the $A_4=0$ gauge the spatial components $A_i(\r,t)$ contain
three $O(3)$ symmetrical structures,

\beq
A_i^a(\r,t) = \ee_{ijk}n_j\frac{1-A(r,t)}{r}+
(\delta_{ai}-n_an_i)\frac{B(r,t)}{r}+n_an_i\frac{C(r,t)}{r},\;\;\;
{\bf n}=\frac{\r}{r} .
\la{O3}\eeq
In Appendix B we quote the relation of these structures to the $A_\mu$ field
in the Lorentz gauge, and the Y--M action written in terms of the $A,B,C$
functions.

In order to find the fields along branches II and III corresponding to
a non-zero field energy $E_{field}$, one has to solve the Y--M
equations (generally speaking, coupled to Higgses) in the time interval
$-T/2<t<T/2$ with the singular boundary conditions \ur{boundary}. As in
the quantum mechanical example, the field energy $E_{field}$ is directly
related to the time interval $T$. However difficult technically, solving
the Y--M equations with given boundary conditions is a well-formulated
problem which can be solved numerically.  We solve it analytically in two
limiting cases:  (i) $\rho\ll T$ corresponding to low energies,
$\alpha E\rho\ll 1$, and (ii) $\rho\gg T$ corresponding to high
energies, $\alpha E\rho\gg 1$.  In the first case we reproduce
the usual instanton results.  The second case corresponds to asymptotically
large energies, and the results here are new.

\subsection{Reproducing Instanton Results}

In the case $\rho \ll T$ , i.e. low energies, we deal with a situation
shown schematically in Fig.1b. Therefore, one has first to find the
singular solutions along branches II and III, then go to the Minkowski
space and find the analogue of the turning point $\phi_3$. Then one has to
construct again the Euclidean solution along branch V. Each time the
boundary conditions for the solutions are determined at the previous step,
so that the procedure is straightforward. However, we have not solved
the problem at arbitrary values of $\rho/T$ but only in the lowest order
in $\rho/T$. The explicit construction of the Y--M fields along the
branches I -- V is rather lengthy and we relegate it to Appendix C.
We find the following expressions for the {\em full} actions along
branches I+IV (with the actions along branches II and III
subtracted -- we call it $S^{I-IV}$, see \eq{SI-IV}) and along the branch V:

\beq
S^{I-IV}_{full}=\frac{4\pi}{\alpha}\left[6\left(\frac{\rho}{T_2}\right)^4
+O\left(\frac{\rho^6}{T_2^6}\right)\right],
\la{SIIVF}\eeq

\beq
S^{V}_{full}=\frac{4\pi}{\alpha}\left[1
-6\left(\frac{\rho}{T_2}\right)^4
-6\left(\frac{\rho}{T_1-T_2}\right)^4
+O\left(\frac{\rho^6}{T_2^6}\right)\right].
\la{SVF}\eeq
In \eqs{SIIVF}{SVF} we have introduced the time $T_2\equiv T-T_1$ which is
the net time for branches II+III.

According to the general mechanical formula (see Appendix A), the energy of
the field is the derivative of the full action in respect to the period of
motion, which is $T_2$ for branches II+III and $T_1$ for branch V. Also,
the field energy $E_{field}$ should be the same on branches II, III and V.
From \eqs{SIIVF}{SVF} we have:

\[
E_{field}=
-\frac{dS_{full}^{I-IV}}{dT_2}=\frac{4\pi}{\alpha}24\frac{\rho^4}{T_2^5}
\]
\beq
=\frac{dS_{full}^{V}}{dT_1}=\frac{4\pi}{\alpha}24\frac{\rho^4}{(T_1-T_2)^5}.  \la{Econs}\eeq

We hence get a relation between the times along branches V and II+III:

\beq
T_1=2T_2, \;\;\;\; {\rm where}\;\;\;\;T_2=\rho\left(\frac{\alpha
E\rho}{96\pi}\right)^{-1/5}.
\la{T1T2}\eeq

In section 2 we have obtained the same relation for the case of quantum
mechanics and presented arguments that it could be of a more
general nature. Now we have demonstarted the validity of this relation
"experimentally" in the Y--M theory, at least for large periods, or,
equivalently, for small energies $E$.

The correspondent shortenned action are (see Appendix A for the
definitions):

\beq
S_{short}^V=S_{full}^{V}-E_{field}T_1=\frac{4\pi}{\alpha}\left[1
-60\left(\frac{\alpha E\rho}{96\pi}\right)^{4/5}
\right],
\la{SshortVlow}\eeq

\beq
S_{short}^{I-IV}=S_{full}^{I-IV}+E_{field}T_2=\frac{4\pi}{\alpha}
30\left(\frac{\alpha E\rho}{96\pi}\right)^{4/5}.
\la{SshortIIVlow}\eeq

Adding these two actions in the exponent in accordance with \eq{2terms}
we get finally the cross section as a function of energy at small values
of $\alpha E\rho$:

\beq
\sigma(E)=\exp\left\{\frac{4\pi}{\alpha}\left[-1
+30\left(\frac{\alpha E\rho}{96\pi}\right)^{4/5}\right]\right\}.
\la{xsectlow}\eeq

This formula coincides with the well-known result for small energies
\cite{Zakharov3,Por} obtained for the instanton-induced total cross
section in the saddle-point approximation. The one-loop correction
to \eq{xsectlow} is also known \cite{KR,WS,Mue2,AM}: it adds
to the exponent of \eq{xsectlow} a term

\beq
-\frac{4\pi}{\alpha}24\left(\frac{\alpha E\rho}{96\pi}\right)^{6/5}.
\la{1looplow}\eeq

The 2-loop correction \cite{DPol} adds

\beq
+\frac{4\pi}{\alpha}\frac{144}{5}\left(\frac{\alpha
E\rho}{96\pi}\right)^{8/5}\left[\ln\frac{96\pi}{\alpha E\rho}
+O(1)\right].
\la{2looplow}\eeq

We have not reproduced these terms directly by our new method (indirectly
we got these terms from the quantum-mechanical calculation of the previous
section by using the conformal-symmetry arguments). We would like to
stress that what looks as a loop expansion in the conventional approach
appears to be equivalent to
finding singular solutions of the {\em classical} eqs. of motion. We
believe that it is a well-formulated program which may be performed for
all values of the dimensional energy $\alpha E\rho$.

\subsection{Singular Solutions for High Energies}

Let us now consider the opposite case: $T\ll\rho$. It means that the
time during which the field is developing is much less than the spatial
spread of the field -- see \eq{boundary}. During this short time the fields
change only at distances $r\ll \rho$ but cannot change significantly at
$r\sim \rho$. Therefore, the region $r>\rho$ cancels out in the
difference between the shortenned actions. For that reason one can neglect
$r$ as compared to $\rho$ in \eq{boundary}, so that the boundary conditions
for the field simplifies.  In terms of the $A,B,C$ functions introduced in
\eq{O3} the boundary condition becomes:

\beq
A\left(r,\pm\frac{T}{2}\right)=3,\;\;B\left(r,\pm\frac{T}{2}\right)=
C\left(r,\pm\frac{T}{2}\right)=
-\frac{2\rho}{r}.
\la{boundsing}\eeq

We see that the structure $B+C$ is singular and thus much larger
than the structures $A$ and $B-C$; therefore we shall neglect the
second two and work with one structure $B+C=D$. The equations
of motion for $A, B-C$ and $D$ will keep $A$ and $B-C$ negligible
as compared to $D$ as far as $r$ will remain much less than $\rho$,
which is the case of interest. Moreover, for the similar reasons one can
neglect the spatial derivatives of $B+C$ and other less singular terms as
compared to the terms $B^4,C^4/r^2$ in the action -- see Appendix B,
\eq{O3A}.  The resulting action written in terms of the large structure
$D$ becomes:

\beq
S^{II+III}=\frac{3}{4\alpha}\int_{-T/2}^{T/2} dt \int_0^\infty dr
\left(\frac{1}{2}\dot{D}^2 + \frac{D^4}{8r^2}\right).
\la{actD}\eeq
The corresponding equation of motion is

\beq
-\ddot{D}+\frac{D^3}{2r^2}=0
\la{eqmD}\eeq
with the boundary contitions

\beq
D\left(r,\pm \frac{T}{2}\right) = -\frac{4\rho}{r}.
\la{boundaryD}\eeq

The problem can be regarded as a simple one-degree-of-freedom mechanical
one, with $r$ viewed as a parameter of the trajectory. \Eq{eqmD} is
integrable as there exists a conserved energy density:

\beq
e(r)=\frac{D^4}{8r^2}-\frac{1}{2}\dot{D}^2.
\la{enedens}\eeq
The time dependence of the field is determined by integrating \eq{enedens}:

\beq
t=\int_{-4\rho/r}^{D(r,t)}\frac{dD}{\sqrt{D^4/4r^2-2e(r)}}.
\la{Doft}\eeq
Introducing dimensionless quantities,

\beq
x=-\frac{D}{[8e(r)r^2]^{1/4}},\;\;\;\;
\beta= \frac{4\rho}{r[8e(r)r^2]^{1/4}},
\la{dimless}\eeq
we rewrite \eq{Doft} as

\beq
\frac{t}{2r}[8e(r)r^2]^{1/4}=\int_x^\beta \frac{dx}{\sqrt{x^4-1}}.
\la{Doftdimless}\eeq
Half-period of the motion, $T/2$, is found from \eq{Doftdimless} as an
integral from the initial configuration $\beta$ up to the turning point
being $x=1$ in the new notations:

\beq
\frac{T}{4r}[8e(r)r^2]^{1/4}=\int_1^\beta \frac{dx}{\sqrt{x^4-1}}.
\la{period}\eeq
It can be written as an equation relating the period and the
energy density:

\beq
\beta\int_1^\beta \frac{dx}{\sqrt{x^4-1}}=\frac{\rho T}{r^2}.
\la{relation}\eeq
This equation has a solution for $\beta$ and hence for $e(r)$ for all
values of the r.h.s.

The shortenned action density along the branches II+III is

\[
s^{II+III}(r)= 2 \int_{-4\rho/r}^{D_{min}} dD
\sqrt{\frac{D^4}{4r^2}-2e(r)}
\]
\beq
= \frac{64\rho^3}{r^4}\frac{1}{\beta^3}\int_1^\beta dx\sqrt{x^4-1}.
\la{actdens}\eeq

This action is apparently divergent, however one should not forget
to subtract a similarly divergent piece corresponding to the zero-energy
branches I and IV. To make the necessary subtraction we notice that
zero energy corresponds to infinite period $T$ and hence to
$\beta\rightarrow\infty$ -- see \eq{relation}. Putting  $\beta=\infty$
in \eq{actdens} we get the shortenned action density for branches I and IV:

\beq
s^{I+IV}(r)= \frac{64\rho^3}{r^4}\frac{1}{3}
\la{actdensI}\eeq
which is indeed independent of $T$, and should be subtracted from
\eq{actdens}. Of course, one could get \eq{actdensI} directly by finding
the action density along the solutions \urs{I}{IV}. To get the complete
shortenned action standing in the exponent for the cross section we
integrate the difference of \ur{actdens} and \ur{actdensI} over $r$. Using
the relation \ur{relation} we integrate over $\beta$ instead of $r$, and
obtain:

\[
S^{I-IV}=\frac{3}{4\alpha}\int_0^\infty\! dr(s^{I+IV}-s^{II+III})=
\frac{4\pi}{\alpha}\left(\frac{\rho}{T}\right)^{\frac{3}{2}}
\frac{6}{\pi}\int_1^\infty\! \frac{d\beta}{\surd\beta}
\left(\int_1^\beta\!\frac{dx}{\sqrt{x^4-1}}\right)^{1/2}
\]
\[
\cdot\left(\int_1^\beta\frac{dx}{\sqrt{x^4-1}}
+ \frac{\beta}{\sqrt{\beta^4-1}}\right)
\left(\frac{1}{3}-\frac{1}{\beta}\int_1^\beta dx \sqrt{x^4-1}\right)
\]
\beq
=\frac{4\pi}{\alpha}5\left(\frac{B(\frac{1}{4},\frac{1}{2})}{8}\right)^4
\left(\frac{\rho}{T}\right)^{3/2}
=\frac{4\pi}{\alpha}0.9232\left(\frac{\rho}{T}\right)^{3/2}.
\la{Acthigh}\eeq
We remind the reader that the above calculation is performed for the
case $\rho \gg T$, hence \eq{Acthigh} describes a rapid fall-off of the
cross section in this region.

Integrating the energy density $e(r)$ over $r$ we get the field energy:

\[
E_{field}=\frac{3}{4\alpha}\int_0^\infty dr e(r) = \frac{4\pi}{\alpha T}
\left(\frac{\rho}{T}\right)^{3/2}\frac{3}{\pi}
\]
\[
\cdot\int_1^\infty \frac{d\beta}{\beta^{5/2}} \left(\int_1^\beta
\frac{dx}{\sqrt{x^4-1}}\right)^{3/2}\left(\int_1^\beta\frac{dx}{\sqrt{x^4-1}}
+ \frac{\beta}{\sqrt{\beta^4-1}}\right)
\]
\beq
=\frac{4\pi}{\alpha}\frac{3}{T}\left(\frac{\rho}{T}\right)^{3/2}
\left(\frac{B(\frac{1}{4},\frac{1}{2})}{8}\right)^4
=\frac{4\pi}{\alpha}\frac{0.5539}{T}\left(\frac{\rho}{T}\right)^{3/2}.
\la{Enhigh}\eeq

One can immediatelly check that \eqs{Acthigh}{Enhigh} satisfy the general
relation between the shortenned action and the energy (see Appendix A):

\beq
\frac{dS_{short}^{I-IV}}{dT}=\frac{dE_{field}}{dT}T.
\la{genrel}\eeq

The energy $E_{field}$ \ur{Enhigh} has actually a double meaning. We have
introduced it as the energy of a singular Euclidean field between the
singularity points. It becomes purely potential energy at the turning point
($t=0$) starting from where the field enters the classically allowed region
and hence develops in real Minkowski time. Therefore, it is also the energy
of the outgoing field or, equivalently, the energy of the {\em final}
multiparticle state.

Simultaneously, it is the energy of the few-particle {\em initial} state.
To see that we notice that the cross section is defined via the imaginary
part of the 4-point Green function of the initial particles, see
\eq{sigma}. The high-energy asymptotics of a Green function in the
background field is determined by the positions of the singularities of
the field. In the case depicted in Fig.1a there is one singularity at
$t=-T/2$ for the amplitude and one singularity at $t=+T/2$ for the
conjugated amplitude. Therefore, the asymptotics of the 4-point Green
function and hence of the cross section is

\[
\exp\left[E_1(-T/2)+E_2(-T/2)-E_1T/2)-E_2T/2-S_{full}(T)\right]
\]
\beq
=\exp\left[-E_{particle}T-S_{full}(T)\right]
\la{asgreen}\eeq
where $E_{particle}=E_1+E_2$ is the energy of the initial 2-particle state
and $S_{full}$ is the full or Lagrange action of the field:

\beq
S_{full}=S_{full}^I+S_{full}^{IV}-S_{full}^{II}-S_{full}^{III}.
\la{Sfullsum}\eeq

On the other hand, the full actions are simply related to the shortenned
actions provided the fields satisfy the equations of motion (see Appendix
A):

\beq
S_{full}=S_{short}^I+S_{short}^{IV}-S_{short}^{II}-S_{short}^{III}
-E_{field}T \equiv S^{I-IV}-E_{field}T
\la{Sfullshort}\eeq
(the quantity $S^{I-IV}$ has been calculated in \eq{Acthigh}). We get
thus from \eq{asgreen}:

\beq
\exp\left[-E_{particle}T+S^{I-IV}+E_{field}T\right].
\la{asshort}\eeq

Integrating over the separation between the singularities $T$ we obtain
the energy conservation law,

\beq
E_{particle}=E_{field}\equiv E
\la{encons}\eeq
whereas the cross section is given simply by

\beq
\sigma(E)=\exp\left[-S^{I-IV}(E)\right]
\la{crosss}\eeq
where the shortenned action $S^{I-IV}$ should be expressed through the
energy. It should be mentioned that in a case depicted in Fig.1b there
are additional field singularities at  $t=\pm T_1/2$ and a similar analysis
is slightly more complicated.

Expressing the time between the singularities $T$ through the field energy
\ur{Enhigh} and substituting it into \eq{Acthigh} we get finally

\beq
\sigma(E)=\exp\left[-\frac{4\pi}{\alpha}1.316
\left(\frac{\alpha\rho E}{4\pi}\right)^{3/5}\right]
\la{xsecthigh}\eeq
which is valid for $\alpha E\rho\gg 1$ and describes a decreasing
cross section at large values of this parameter.

The result of this subsection can be formulated also in terms
of the instan\-ton--anti-instanton interaction potential, \ui.
We find at small separations $T$ a strong repulsion,

\beq
U_{int}(T)=S_{full}^{I-IV}(T)=S^{I-IV}-ET=\frac{4\pi}{\alpha}0.3693
\left(\frac{\rho}{T}\right)^{3/2},\;\;\;\;T\ll\rho,
\la{repul}\eeq
as contrasted to the usual attraction at large separations, see
\eq{uintym}. We believe that it is a reasonable result, if one
defines \ui not as the action of an arbitrarily chosen Euclidean
configuration like the valley (we have shown in section 2 that the
valley leads to the wrong \ui already in the next-to-leading order),
but through unitarity, as in \eq{uint}. Physically, the effective
repulsion of instantons at small separations appears here as a result
of a rapidly decreasing overlap between low-energy and high-energy states.
From the purely Euclidean viewpoint a closely situated instanton and
anti-intstanton resemble a vacuum state and therefore have to be
effectively strongly repulsive if we wish to read off a
non-perturbative contribution from that configuration. \Eq{repul}
is good news for the QCD instanton-vacuum builders.

\section{Energy Behaviour of the Cross Section}
\setcounter{equation}{0}
\def\theequation{\arabic{section}.\arabic{equation}}

We have established the behaviour of the cross section for
multi-gauge-boson production as a function of a dimensionless
quantity $\alpha E \rho$ where $\rho$ is the spatial size of the
Y--M field, at small (\eq{xsectlow}) and large
(\eq{xsecthigh}) values of this parameter. Applications of this
result are different in the QCD case where scale invariance is preserved
at the classical level, and in the electroweak theory where it is
broken from the beginning. We discuss the two theories in turn.

\subsection{QCD}

In QCD the scale $\rho$ is arbitrary, and one has to integrate over
it. Naturally, $\rho$ will be then fixed by the maximum of the
cross section as function of the dimensionless parameter $\alpha E\rho$.
We know the behaviour of the $\log \sigma(E)$ at small and large values
of this parameter, and we plot it in Fig.3, where we have added the
next-to-leading term \ur{1looplow} at small $\alpha E\rho$. The low-energy
and high-energy curves intersect and give the maximum at a point which is
remarkably close to a point marked by a circle in Fig.3, which is the
position of the maximum that we would expect independently of the
above calculations.

Indeed, in Section 2 we have presented general arguments in
favour of the relation \ur{V-I}, and in Section 3 we have confirmed it
at low energies for the Y--M case. If we assume \eq{V-I} to be correct
at all energies up to the sphaleron top of the barrier, we would get
that at the maximum corresponding to the top of the barrier the cross
section is exactly the square root of the cross section at zero energy.
It means that the low- and high-energy branches of the $\log$ of the
cross section should intersect at

\beq
\log \sigma_{max}=-\frac{4\pi}{\alpha}\frac{1}{2}.
\la{xsymmax}\eeq

To what value of the dimensionless parameter $\alpha E\rho$ does this
maximum correspond? Our arguments here are even more shaky, however,
one could speculate that the maximum occurs at energy equal to the
mass of the sphaleron. In the massless Y--M theory we are now dealing with,
the sphaleron mass can be estimated as the potential energy of the
instanton field exactly in the middle of the transition when the
instanton passes the $N_{CS}=1/2$ point. We have:

\beq
M_{sph}(\rho)=\frac{1}{4g^2}4\pi \int dr r^2\left.\frac{96\rho^4}
{(t^2+r^2+\rho^2)^4}\right|_{t=o}
=\frac{4\pi}{\alpha}\frac{1}{\rho}\frac{3}{16}.
\la{sphal}\eeq

We find then that the maximum of the cross section should be achieved
at the value of the dimensionless parameter

\beq
\left(\frac{\alpha E\rho}{4\pi}\right)_{max}=\frac{3}{16}=0.1875.
\la{maxEr}\eeq
This is the point marked by a circle in Fig.3, and we see that it is
remarkably close to the intersection of the low- and high-energy curves.
Higher order corrections to both curves should probably move the
intersection point exactly to the position calculated here.

The fact that the spatial size of the classical field $\rho$ scales
down as $1/E$ at the maximum means that quantum corrections to the
classical calculations of this paper should be controlable and small
at $E\rightarrow\infty$. The one-loop quantum corrections, as usually,
should make the gauge coupling run. Also, one would expect a large
prefactor owing to the zero modes about the classical trajectory --
something like $(2\pi/\alpha)^{4N_c}$, where $N_c$ is the number of colours.
The prefactor $\alpha$ starts to "run" only at the 2-loop
level. Combining renormalization-group arguments with the result
\ur{xsymmax} we expect the isotropical multi-gluon production cross section
to be

\beq
\sigma_{jet}(E)=\frac{c}{E^2}\left(\frac{\Lambda}{E}\right)^b
\left(b\ln\frac{E}{\Lambda}\right)^{4N_c+p}
\left[1+O\left(\frac{1}{\ln(E/\Lambda)}\right)\right]
\la{jetprod}\eeq
where $b=11N_c/3-2N_f/3\approx 7$ is the 1-loop Gell-Mann--Low coefficient
and $\Lambda$ is $\Lambda_{QCD}\approx 200 MeV$.
The coefficient $c$ and the power $p$ are determined by the second loop
Gell-Mann--Low coefficient and by the Green functions of the initial
particles of the process; that is where the non-universality of cross
sections comes in. A calculation of these constants seems to be
feasible.

We thus arrive to a prediction that the total cross section for
isotropical multi-gluon production in QCD should fall off as
approximately the ninth power of their aggregate energy. However a
presumably large prefactor and the peculiarity of the events might help to
make such processes observable; this subject deserves a more thorough
study.

It is worth mentioning that the momentum distribution of the produced
multi-gluon state can be easily found by solving the Y--M eqs. in Minkowski
time starting from the "sphaleron" field \ur{sphal}. Since that field has
only one parameter $\rho$ related to energy through \eq{maxEr},
the gluon momentum distrubution will have a scaling behaviour with energy,
modulo logarithmic corrections.

\subsection{EW Theory}

In this case the size of the classical field $\rho$ is determined not only
by energy but also by the Higgs v.e.v. $v$. At low energies one has to
multiply the $\alpha E\rho$-dependent cross section \ur{xsectlow}
(with the addition of 1-loop \ur{1looplow} and 2-loop \ur{2looplow}
corrections) by the factor

\beq
\exp(-2\pi^2\rho^2v^2).
\la{Hfactor}\eeq
Integrating over $\rho$ one gets the familiar expansion of the total baryon
number violating (BNV) cross section in terms of the dimensionless energy
measured in units of the sphaleron mass \cite{DPol}:

\[
\sigma_{tot}^{\not{B}}(E)=\exp\left[\frac{4\pi}{\alpha}F(\e)\right],
\;\;\;\;\e=\frac{E}{\surd 6\pi m_W/\alpha},
\]
\beq
F(\e)=-1+\frac{9}{8}\e^{4/3}-\frac{9}{16}\e^{6/3}-
\frac{3}{16}\e^{8/3}\left[\ln\e+O(1)\right].
\la{fe}\eeq

The growth of the function $F(\e)$ has lead in the past to
hopes that the BNV processes may become unsuppressed at energies
of the order of the sphaleron mass, $m_W/\alpha$, corresponding to
$\e\sim 1$. However, unitarity arguments of refs.\cite{Zakharov2,MS,V}
indicated that the BNV cross section induced by instantons should not
rise beyond the square root of the cross section at zero energies. We
now arrive to the same conclusion from our singular solutions.

At very high energies ($E\gg m_W/\alpha$) the size of the field $\rho$
is determined by the factor \ur{xsecthigh} rather than by the Higgs factor
like \ur{Hfactor}, meaning that $\rho \sim 1/\alpha E \ll 1/m_W$. Therefore
one can neglect the Higgs influence and arrive to the conclusions of the
previous subsection, viz. that the cross section does not rise above the
value of $\exp(-2\pi/\alpha)\approx 10^{-85}$.

At energies of the order of the sphaleron mass ($E\sim m_w/\alpha$) one
cannot neglect the Higgs effects and has to find singular sloutions of
the Y--M--H equations along the branches I-V. If our general arguments
based on analyticity are correct, we would again expect that the maximum
occurs at exactly the sphaleron energy and corresponds to the square root
suppression.

The momentum distribution of the produced $W$ and $H$ bosons should
follow from the Minkowski development of the sphaleron -- the "fall"
of the sphaleron. This distribution has been studied
numerically in refs. \cite{HK,CH}. We argue now that the same distribution
would happen at high-energy collisions, however its probability would
be of the order of $10^{-85}$.

\section{Massless $\lambda \phi^4$ Theory}
\setcounter{equation}{0}
\def\theequation{\arabic{section}.\arabic{equation}}

For future reference we would like to cite here the results for the
massless $\lambda\phi^4$ theory whose Euclidean action is

\beq
S=\int\! d^4x \left( \frac{1}{2} (\partial_\mu
\phi)^2+\frac{\lambda}{4}\phi^4 \right)
\la{actphi}\eeq

Instead of the singular instanton one has to start here with the singular
Lipaton \cite{Lip} which gives an $O(3)$ symmetrical
zero-energy field singular at $t=\pm T/2$ \cite{Khlebnikov}:

\beq
\phi(\r,t)=2\sqrt\frac{2}{\lambda}\cdot
\frac{\rho}{r^2+(t\pm T/2\mp\rho)^2-\rho^2}.
\la{lip}\eeq

Since there is no tunneling and consequently no branch V in this
theory, one gets a monotonically decreasing cross section as function
of the dimensionless quantity $\lambda E \rho$. Probably it is in
accordance with the theory being not asymptotically free.
Calculations of Section 3 can be repeated without any serious change
(in fact they are much more simple). For small $\lambda E\rho$ the cross
section has been actually calculated in ref.\cite{Khlebnikov}, and we
confirm it:

\beq
\log\sigma_{tot}=-\frac{16\pi^2}{\lambda}\frac{3}{2^{1/3}}\left(
\frac{E\rho\lambda}{16\pi^2}\right)^{2/3},\;\;\;\;\lambda E\rho\ll 1.
\la{philow}\eeq

For large $\lambda E\rho$ the leading term can be obtained immediately from
\eq{xsecthigh} by changing notations. Indeed, introducing a new variable
$D(r,t)$,

\beq
\phi(r,t)=\frac{1}{\sqrt{2\lambda}}\frac{D(r,t)}{r},
\la{newvarD}\eeq
we get for the action \ur{actphi}:

\beq
S=\frac{2\pi}{\lambda}\int_{-T/2}^{T/2}\! dt \int \! dr
\left(\frac{\dot{D}^2}{2} +\frac{D^4}{8r^2}\right)
\la{actDphi}\eeq
which coincides with \ur{actD} up to the overall factor.
The boundary conditions for $D(r,t)$ as imposed by \eq{lip} also coincide
with those of the Y--M case \ur{boundaryD}. Hence the behaviour of the
cross section can be immediatelly obtained from \eq{xsecthigh} by a
suitable change of constants. We get:

\beq
\log \sigma_{tot}=-\frac{16\pi^2}{\lambda}1.119
\left(\frac{\lambda E\rho}{16\pi^2} \right)^{3/5}, \;\;\;\;
\lambda E\rho\gg 1.
\la{phihigh}\eeq

\Eqs{philow}{phihigh} exhibit a monotonically decreasing cross section
as function of $\lambda E\rho$; integrating over $\rho$ one falls into
the trivial maximum at $\lambda E\rho=0$ corresponding to zero particle
production, which is naturally more probable than multi-particle
production. It is a prerogative of field theories with non-trivial
topology to have a non-trivial behaviour of classical cross sections even
in the massless limit.

\section{Conclusions}

We have elaborated the use of singular Euclidean solutions to describe
semi-classical multi-particle production in field theories. We have shown
that using instanton-like $O(3)$ symmetrical singular solutions one
reproduces the usual formulae for the instanton-induced cross sections
for low energies. However the present approach allows to solve the problem
at all energies. We have found analytically a new family of singular
solutions labelled by $\rho/T$, which govern the high-energy behaviour
of multi-particle production -- both for the Y--M and $\lambda\phi^4$
theories. We have presented new arguments (not based on unitarity but
rather on analyticity) that the maximum of the instanton-induced cross
sections in QCD and EW theories is still exponentially small in the
coupling constant, with the coefficient corresponding exactly to the
"square root" suppression as compared to zero energy.  An interpolation of
the low- and high-energy branches of the cross section give numerically
this "square root" point to a surprisingly good accuracy -- see Fig.3.
Simultaneously our result means that there is effectively a strong
repulsion between instantons at small separations.

As a by-product of our study we get a non-perturbative formula for the
isotropic multi-gluon production in QCD, \eq{jetprod}. Presently unknown
coefficients and powers of the logarithms in that formula seem to be
calculable, and deserve further study. It would be also extremely useful,
if possible, to understand \eq{jetprod} in terms of some clever summation
of Feynman graphs for a gluon branching process -- we mean the approach
initiated by Cornwall \cite{Corn} and Goldberg \cite{Gold}.

As to the applications to the baryon number violating processes in the EW
theory, our result is in the line with other people who in the recent
years argued from various angles that the BNV instanton-induced cross
section has to be small at all energies. However we feel that it is not
the end of the story but, rather, its beginning. It should not be surprising
that the {\em isotropic} multi-particle production is suppressed at high
energies. One should expect that at high energies the events have at most
the axial symmetry.  Therefore, one should look for solutions with
singularities of a different kind -- not just $O(3)$ symmetrical. The same
applies to other cases where we expect that multi-particle production might
be described semi-classically, -- such as multi-jet production and low-x
structure functions in QCD, multi-pion production in heavy-ion collisions,
etc.

Our final remark concerns an alternative approach to semi-classical
multi-particle production \cite{Larry} in which the few initial
high-energy particles are used as a (weak) source for classical fields. How
can this approach be reconciled with our? We think that including the
source explicitly, one has to look for anomalous solutions of field
equations, which do not possess singularities whatever small is the source
but which develop a singularity when the source is put identically
to zero. We have observed such scenario to happen in quantum mechanics and
a d=2 massive $\sigma$ model, and believe that it is a general case.
However, in a d=4 Yang--Mills theory the hopefully equivalent approach
based on singular Euclidean solutions is seemingly more simple.

\section{Acknowledgements}

We are grateful to Larry McLerran for an enlighting discussion of the
Mueller corrections. We would like to thank the Institute for Theoretical
Physics-II of the Ruhr University at Bochum for hospitality. D.D.
acknowledges the support of the A.v.Humboldt - Stiftung.

\vskip 1cm

\appendix

\section{A. ~General Relations for the Action}
\setcounter{equation}{0}
\def\theequation{\Alph{section}.\arabic{equation}}

Let $S_{full}$ be the Lagrange action:

\beq
S_{full}=\frac{1}{g^2}\int dt \left[\frac{1}{2}\dot{\phi}^2+U(\phi)\right].
\la{Sfull}\eeq
Using the energy conservation,

\beq
\frac{1}{2}\dot{\phi}^2-U(\phi)=-Eg^2,
\la{enconse}\eeq
which is fulfilled if the field $\phi(t)$ satisfies the eqs. of motion,
we can present $S_{full}$ as

\[
S_{full}=\frac{1}{g^2}\int dt \dot{\phi}^2 + E\int dt =
\]
\beq
\frac{1}{2g^2}\int d\phi \sqrt{U(\phi)-\e}+ET=S_{short}+ET
\la{Sshort}\eeq
where $T$ is the period of the motion. \Eq{Sshort} introduces the so-called
shortenned action,

\beq
S_{short}=\int pdq.
\la{pq}\eeq

In the main text we make intensive use of the general identities
following from \eq{Sshort}:

\beq
\frac{dS_{short}}{dE}= -T, \;\;\;\; \frac{dS_{short}}{dT}= -T\frac{dE}{dT}
\la{id1}\eeq
and

\beq
\frac{dS_{full}}{dT}= E.
\la{id2}\eeq

Branches I and IV correspond to zero energy, so that the full actions
along these branches coincide with the shortenned ones:

\beq
S_{full}^I=S_{full}^{IV}=S_{short}^I=S_{short}^{IV}.
\la{fsI}\eeq

The relations \urs{id1}{id2} apply directly to branch V in which case
the period is denoted by $T_1$, see Fig.1b:

\beq
\frac{dS_{short}^V}{dE}= -T_1, \;\;\;\; \frac{dS_{short}^V}{dT_1}=
-T_1\frac{dE}{dT_1},\;\;\;\;\frac{dS_{full}^V}{dT_1}= E.
\la{idV}\eeq

The period of motion along branches II+III is denoted by $T_2$ (it
becomes equal to the full time between the singularities $T$ for
energies above the barrier), and we have:

\beq
S_{full}^{II+III}=S_{short}^{II+III}+ET_2
\la{fsII}\eeq

We usually combine this action with that along branches I and IV in
order to cancel the divergencies. Denoting

\beq
S_{full}^{I-IV}\equiv S_{full}^{I+IV}-S_{full}^{II+III},\;\;\;\;
S^{I-IV}\equiv S_{short}^{I+IV}-S_{short}^{II+III},
\la{IIV}\eeq
we have

\beq
S_{full}^{I-IV}=S^{I-IV}-ET_2
\la{fsIIV}\eeq
so that

\beq
\frac{dS^{I-IV}}{dE}= T_2, \;\;\;\; \frac{dS^{I-IV}}{dT_2}=
T_2\frac{dE}{dT_2},\;\;\;\; \frac{dS_{full}^{I-IV}}{dT_2}=-E.
\la{idIIV}\eeq

\section{B. Gauge transformation to the $A_4=0$ Gauge}
\setcounter{equation}{0}
\def\theequation{\Alph{section}.\arabic{equation}}

Let us consider an instanton-like field in the Lorentz gauge:

\beq
A_\mu^a(x)=2\bar{\eta}_{\mu\nu}^a x_\nu\Phi(x^2)
\la{Lorentz}\eeq
or

\[
A_4=A_4^a\frac{\tau^a}{2}=({\bf n\tau})r\Phi(r^2+t^2),
\]
\beq
A_i=A_i^a\frac{\tau^a}{2}=\left[-\tau_i t+i\left(({\bf
n\tau})\tau_i-n_i\right)r\right]\Phi(r^2+t^2).
\la{compon}\eeq

We make a hedgehog gauge transformation

\beq
A_\mu^\prime=U^\dagger(\r,t) A_\mu U(\r,t)
+iU^\dagger\partial_\mu U, \;\;\; U=\exp\left[i({\bf n\tau})P(r,t)\right]
\la{GTH}\eeq
in order to eliminate the $A_4^\prime$ component. The profile function of
the gauge transformation is found from the equation

\beq
A_4=iU\partial_4U^\dagger=({\bf n\tau})\frac{\partial P}{\partial t}=
({\bf n\tau})r\Phi(r^2+t^2)
\la{profdif}\eeq
which can be integrated once the function $\Phi(r^2+t^2)$ is given.

With $P(r,t)$ known, the gauge-transformed spatial components $A_i^\prime$
take the $O(3)$ symmetrical form,

\beq
A_i^{\prime a}(\r,t) = \ee_{ijk}n_j\frac{1-A(r,t)}{r}+
(\delta_{ai}-n_an_i)\frac{B(r,t)}{r}+n_an_i\frac{C(r,t)}{r},\;\;\;
{\bf n}=\frac{\r}{r}
\la{O3app}\eeq
where

\beq
A(r,t)=\cos 2P+(\cos 2P r +\sin 2P t)2r\Phi,
\la{A}\eeq

\beq
B(r,t)=-\sin 2P+(\sin 2P r -\cos 2P t)2r\Phi,
\la{B}\eeq

\beq
C(r,t)=-2r\left[\frac{\partial P}{\partial r}+t\Phi\right].
\la{C}\eeq

If the 't Hooft symbol $\eta$ instead of $\bar{\eta}$ is used in
the definition of the field \ur{Lorentz}, one has to change
$t\rightarrow -t$ in all the above equations.

The Y--M action rewritten in terms of these structures is

\[
S_{Y-M}=\frac{1}{4g^2}\int d^4x \left(F_{\mu\nu}^a\right)^2=
\frac{1}{\alpha} \int dt \int dr\left[\dot{A}^2+\dot{B}^2
+\frac{1}{2}\dot{C}^2\right.
\]
\beq
+\left. A^{\prime 2}+B^{\prime 2}+\frac{(A^2+B^2-1)^2}{2r^2}+
\frac{2C(A^\prime B-AB^\prime)}{r}+\frac{C^2(A^2+B^2)}{r^2}\right].
\la{O3A}\eeq

\section{C. Singular Y--M fields at Low Energies}
\setcounter{equation}{0}
\def\theequation{\Alph{section}.\arabic{equation}}
 \def\thesubsection{\Alph{section}.\arabic{subsection}}
\subsection{Branches II and III}

At low energies we do not expect large deviation on branches II and
III from the zero-energy solutions \urs{I}{IV} which are just singular
instantons. To save the space
we introduce the following notations. Let $A_\mu^{inst}(t+T/2-\rho)$ be a
singular instanton in the Lorentz gauge defined by \eq{I}. Similarly,
$A_\mu^{anti}(t-T/2+\rho)$ is a short-hand notation for the singular
anti-intstanton given by \eq{IV}. The time arguments of the fields are
what we wish to stress here. We next gauge transform the
fields to the $A_4=0$ gauge according to the formulae of the Appendix B.
We denote the resulting spatial components as $A_i(t+T/2-\rho)$ and
$A_i(-t+T/2-\rho)$ (note that we have changed the sign of the argument
for the case of the anti-instanton field in agreement with the procedure of
the Appendix B).

We next notice that the net time for branches II and III is $T_2\equiv
T-T_1$ (see Fig.1b). The field along branch II has to stop at a turning
point at time $-T_1/2$, and the field along branch III has to stop at time
$T_1/2$.  Since these are the {\em same} turning points, we can temporaly
identify them with a single point $t=0$; the singularities occur then at
time $\pm T_2/2$.

To get the fields satisfying the boundary conditions at the singularity
points we introduce the field

\beq
A_i^0=\left\{\begin{array}{c} A_i(t+T_2/2+\rho),\;\;\;t<0 \\
A_i(-t+T_2/2+\rho),\;\;\;t>0.\end{array}\right.
\la{A0}\eeq

The two branches match at $t=0$, however their time derivatives there
have opposite signs. \Eq{A0} satisfies the Y--M equation of
motion everywhere except the point $t=0$:

\beq
\frac{\delta S[A_i^0]}{\delta A_i}=2A_i(T_2/2+\rho)\delta(t)=
\delta(t)O(\rho^2/T_2^2).
\la{eqmoz}\eeq

Therefore, we have to modify $A^0$ in the vicinity of the turning point
at $t=0$. We look for the solutions along the branches II and III in the
form

\beq
A_i(t)=A_i^0(t)+B_i(t)
\la{II}\eeq
where the additional field $B_i$ satisfies the zero boundary
conditions at the singularity points, $\pm T_2/2$, and the equation of
motion:

\beq
0=\frac{\delta S[A_i^0+B_i]}{\delta A_i}=\frac{\delta S[A_i^0]}{\delta A_i}
+(-\partial_t^2+{\cal K}(A_i^0))B+O(B^2)
\la{eqmoB}\eeq
where $-\partial_t^2+{\cal K}$ is the quadratic form of the Y--M action.

To the accuracy we are now interested in, the field $B_i$ is of the order
of $\rho^2/T_2^2$. Therefore, we can neglect the non-linear terms in
\eq{eqmoB} and the shifts by $\rho$ in the arguments of the fields.
Furthermore, in the region of $t\approx 0$ one can put $A^0=0$ in the
quadratic form ${\cal K}$, since $A^0$ is also of the order of
$\rho^2/T_2^2$ in this region.  Solving \eq{eqmoB} at $t\approx 0$
and taking into account \eq{eqmoz} we get:

\beq
B_i(t\approx 0)=A_i(|t|+T_2/2).
\la{Bt}\eeq

The resulting field $A_i^0+B_i$ is smooth at $t=0$; it is shown
schematically in Fig.4a. \Eq{Bt} gives the only solution of the equation of
motion  to the needed accuracy, which is even in $t$ and decreases in the
direction of the singularity points. One could, in principle, add to
\eq{Bt} a solution of the homogeneous equation. Such solutions would be the
zero modes of the singular instanton; their addition would correspond to a
shift or a distortion of the singularity points and therefore would not
satisfy the necessary boundary conditions at $\pm T_2/2$.

We thus obtain the field at the turning point $t=0$:

\beq
\dot{A}_i^{II,III}(\r,0) = 0, \;\;\;\;
A_i^{II,III}(\r,0)=2A_i(\r,T_2/2),
\la{AII}\eeq

Let us now calculate the full action along the branches II and III,
subtracting those along the zero-energy branches I and IV:

\beq
S^{I-IV}_{full}=S^{I+IV}-S^{II+III}_{full}=
2\int_{-\infty}^{-T_2/2}\! dt\int\! d^3x {\cal L}[A^{I}]
-\int_{-T_2/2}^{T_2/2}\! dt \int\! d^3x {\cal L}[A^{II,III}].
\la{defact}\eeq

Using equations of motion and gauge invariance this action can be
calculated without undue difficulty. Let us add and subtract the action
computed on $A^0$ \ur{A0} in the time interval between $\mp T_2/2$. We
write:

\[
S^{I-IV}_{full}=W_1+W_2,
\]
\[
W_1=2\int_{-\infty}^{-T_2/2} dt \int d^3x {\cal L}[A_i(t+T_2/2)]
\]
\[
-2\int_{-T_2/2}^0 dt\int d^3x{\cal L}[A_i(t+T_2/2)],
\]
\beq
W_2=-\int_{-T_2/2}^{T_2/2} dt\int d^3x\left\{{\cal L}[A^0+B]-
{\cal L}[A^0]\right\}.
\la{twoparts}\eeq

In $W_1$ we change the time variables and use the fact that the Lagrange
density is even in time. We obtain:

\beq
W_1=2\int_{T_2/2}^{\infty}dt\int d^3x{\cal L}[A_i(t)].
\la{W1}\eeq

This integral can be calculated in two ways: (i) using the equation
of motion for the $A_i$ field and (ii) using the gauge invariance. The
first method leads to the equation:

\beq
W_1=-\frac{1}{g^2}\int d^3x A_i^a(\r,T_2/2) \dot{A}_i^a(\r,T_2/2).
\la{W11}\eeq

Using the gauge invariance we can return to the Lorentz gauge and obtain:

\[
W_1=\frac{1}{2g^2}\int_{T_2/2}^{\infty}dt \int d^3x
\left(F_{\mu\nu}^a\right)^2
\]
\beq
=\frac{4\pi}{\alpha}\left[\left(\frac{\rho}{T_2}\right)^4
+O\left(\frac{\rho^6}{T_2^6}\right)\right].
\la{W12}\eeq

In $W_2$ we expand the integrand in $B_i$ neglecting terms $O(B^3)$ and
$O(B^4)$ and use the equations of motion for $A_i^0$ and $B_i$
(\eqs{eqmoz}{eqmoB}). We have:

\[
W_2=-\frac{1}{g^2}\int_{-T_2/2}^{T_2/2} dt \int d^3x B_i^a(\r,t)
\dot{A}_i^a(\r,T_2)\delta(t)
\]
\beq
=-\frac{1}{g^2}\int d^3x A_i^a(\r,T_2/2) \dot{A}_i^a(\r,T_2/2)=W_1
\la{W2cont}\eeq
where \eq{W11} has been used in the last line. Adding up the two pieces we
thus obtain the full action along branches II+III (with branches I
and IV subtracted):

\beq
S^{I-IV}_{full}=\frac{4\pi}{\alpha}\left[6\left(\frac{\rho}{T_2}\right)^4
+O\left(\frac{\rho^6}{T_2^6}\right)\right].
\la{SIIVFA}\eeq

\subsection{Branch V}

We now go back to the turning point at $t=0$ where we have already found
the field -- it is given by \eq{AII}. What we called $t=0$ corresponds
actually to two points:  $t=\pm T_1/2$. At these points the field enters
the allowed region and develops in Minkowski time. The solution of the
Minkowski equations of motion is

\beq
A_i^{Mink}=A_i(T_2/2+it)+A_i(T_2/2-it).
\la{Mink}\eeq
Indeed, this is a solution of Minkowski eqs. of motion for not too large
$t$ since $A_i$ itself is a solution of Euclidean ones, and \ur{Mink} is
evidently obeying boundary conditions \ur{AII}.

The field \ur{Mink} describes the outgoing gluon field which corresponds to
multi-gluon production without changing of the Chern--Simons
number. In order to change it we need one more tunneling transition
corresponding to branch $V$  of Fig.1b. Normally the Minkowski solution
develops between the turning points $\phi_1$ and $\phi_3$ (as it was in
our quantum mechanical example), and determines the initial field
configuration $\phi_3$ for branch V. However this is not the case in the
lowest approximation we are dealing with. In this approximation the two
turning points, $\phi_3$ and $\phi_1$, coincide, so that branch V starts at
the same field where branch II ends.
It should be also mentioned that far from the centers in the leading
approximation there is no difference between the field $A_i$ of the
singular and of the non-singular instantons. We shall therefore denote the
instanton field in the $A_4=0$ gauge by $A_i$ as before.

We have thus to find branch V as a solution of the Euclidean eqs. of
motion with the boundary conditions given by \eq{AII}:

\beq
\dot{A}_i^{V}(\pm T_1/2) = 0, \;\;\;\; A_i^{V}(\pm T_1/2)=
2A_i(T_2/2).
\la{AV}\eeq

Again we are looking for a solution in the form

\beq
A_i^V=A^0_i+B_i
\la{Vgen}\eeq
where this time $A_i^0$ is the field of the usual instanton in the
$A_4=0$ gauge at $t<0$ and that of an anti-instanton at $t>0$; the
centers of both are shifted by $\Delta T$, to be found below:

\beq
A_i^0=\left\{\begin{array}{c} A_i(t+T_1/2-\Delta T),\;\;\;t<0 \\
A_i(-t+T_2/2-\Delta T),\;\;\;t>0.\end{array}\right.
\la{A00}\eeq
(see Fig.4b).

The shift in the positions of the instantons, $\Delta T$, should be
determined from the region near $-T_1/2$. In this region the field $B_i$
obeys linearized eqs. of motion following from the quadratic form of the
action in which one can neglect the instanton field being small in this
region.  The boundary conditions for $B_i$ follow directly from
\eqs{AV}{A00}:

\beq
B_i(t=-T_1/2)=2 A_i(T_2/2)-A_i(\Delta T),\;\;\;\;
\dot{B}_i(t=-T_1/2)=-\dot{A}_i(\Delta T).
\la{bouB}\eeq

The field $B_i$ should not increase in the direction to
the center of instantons -- then it does not include zero modes
corresponding to the distortion of the instantons. We can fulfill
this requirement only if $\Delta T=T_2/2$. Indeed, in this case the
solution of eq. of motion for $B_i$ with the boundary conditions
\ur{bouB} is

\beq
B_i=A_i\left(-t+(T_1-T_2)/2\right),
\la{Bsol}\eeq
which clearly decreases for increasing $t$ and hence does not contain
the zero modes which would lead to the shift of the instanton position.

Let us turn now to the region near $t=0$. We cannot apply here immediately
the method used above for branches II and III since the field $A_i^0$
is not small now. However it is {\em gauge equivalent} to a small field
$\tilde{A}_i^0$:

\beq
A_i^0(t)=U^\dagger(\r)\tilde{A}_i^0 U(\r)+iU^\dagger\partial_iU
\eeq
where $U(\r)$ is a gauge transformation with the winding number equal to
unity. One can apply the same considerations as above to the gauge
transformed fields $\tilde{A}_i^0$ and $\tilde{B}_i=UB_iU^\dagger$.
As a result we obtain the following expression for $B_i$ near $t=0$ (cf.
\eq{Bt}):

\beq
B_i=U^\dagger(\r)\tilde{A}_i\left(|t|+\frac{T_1-T_2}{2}\right)U(\r).
\la{Bfin}\eeq

The solution of Minkowski eqs. of motion, which starts at the
turning point at $t=0$, is (cf. \eq{Mink}):

\beq
A_i^{Mink}(t)=U^\dagger(\r)\left[\tilde{A}_i\left(it+\frac{T_1-T_2}{2}\right)
+\tilde{A}_i\left(-it+\frac{T_1-T_2}{2}\right)
\right]U(\r)+iU^\dagger\partial_i U.
\la{Aminkf}\eeq
It describes gluon production around the minimum with Chern-Simons number
equal to one.

Now we can calculate the action along the branch V. The full action,

\beq
S_{full}^V=\frac{2}{g^2}\int_{-T_1/2}^0\int d^3x{\cal L}[A^0_i+B_i],
\eeq
can be expressed as a sum of twice the instanton action,

\beq
S^{I\bar{I}}=\frac{2}{g^2}\int_{-\infty}^{\infty}\int d^3x
{\cal L}[A_i]=\frac{4\pi}{\alpha},
\eeq
and corrections of four types. The first is due to the fact that we have
to calculate the action not along the whole $t$ axis but in the time
interval $(-\infty,\,-T_1/2)$:

\beq
\Delta S_1=
-\frac{2}{g^2}\int_{-\infty}^{T_1/2}\int d^3x {\cal L}
\left[A_i\left(t+\frac{T_1-T_2}{2}\right)\right]=
-\frac{4\pi}{\alpha}3\left(\frac{\rho}{T_2}\right)^4.
\la{Delta1}\eeq
The second is a similar defect of the action but coming from the time
interval $(0,\,\infty)$:

\beq
\Delta S_2=-\frac{2}{g^2}\int_{0}^{\infty}\int d^3x
{\cal L}\left[A_i\left(t+\frac{T_1-T_2}{2}\right)\right]=
-\frac{4\pi}{\alpha}3\left(\frac{\rho}{T_1-T_2}\right)^4.
\la{Delta2}\eeq
The third and fourth corrections are the contributions due to the $B_i$
field:

\beq
\Delta S_3+\Delta S_4=\frac{1}{g^2}\int_{-T_1/2}^{T_1/2}\int
d^3x \left\{{\cal L}[A_i^0+B_i]-{\cal L}[A_i^0]\right\}.
\la{Delta3}\eeq

As above, we shall use here the eqs. of motion both for the field
$A^{0}_i$ and $B_i$, and get two contributions. One is the
contribution of the boundaries at $t=\pm T_1/2$:

\beq
\Delta S_3=\frac{2}{g^2}\int d^3x\left[
B_i(t)\dot{A}_i\left(t+\frac{T_1-T_2}{2}\right)+\frac{1}{2}
B_i(t)\dot{B}_i(t) \right]_{t=-T_1/2}
=\Delta S_1.
\la{Delta4}\eeq
The other is the contribution of the $\delta$ function at $t=0$:

\beq
\Delta S_4=\frac{1}{g^2}\int d^3x \left[
B_i(t)\dot{A}_i\left(t+\frac{T_1-T_2}{2}\right)+\frac{1}{2}
B_i(t)\dot{B}_i(t) \right]_{t=0} =\Delta S_2.
\la{Delta5}\eeq

Adding up the four corrections we obtain finally the full action
along branch V:

\beq
S^{V}_{full}=\frac{4\pi}{\alpha}\left[1
-6\left(\frac{\rho}{T_2}\right)^4
-6\left(\frac{\rho}{T_1-T_2}\right)^4\right].
\la{SVFA}\eeq

\newpage
{\Large Figure Captions} \\

Fig.1. Singular trajectories below (a) and above (b) the sphaleron \\

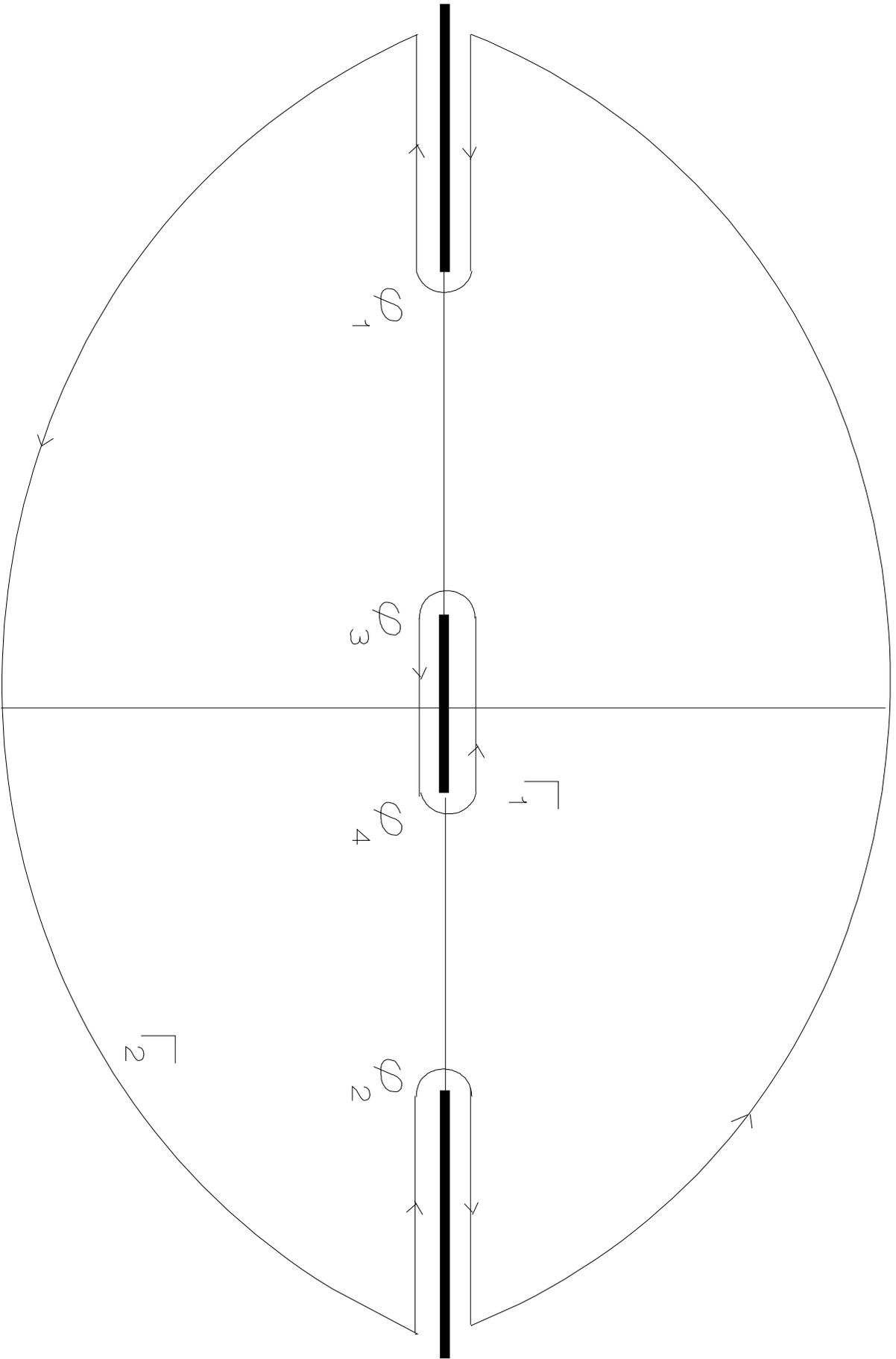
Fig.2. Contour integration proving the relation between the pe\-ri\-ods,
\eq{relper}  \\

Fig.3. The behaviour of the log of the cross section as function of
$\alpha E\rho$ at small (dashed curve) and large (dotted curve) values
of this parameter. The circle shows the expected maximum of the cross
section  \\

Fig.4. A schematic view of the singular solutions along branches II and
III (a) and the bounce solution along branch V (b) \\

\end{document}